\documentclass[twoside,a4paper]{article} 
\usepackage{graphicx}
\usepackage{epsfig}
\usepackage{subfigure}
\begin{document}

\title{Measurement of $\Delta$ and $K^{*}$ Production in $d$+Au collisions 
at $\sqrt{s_{NN}}$ = 200 GeV. }
\author{Dipak Mishra$^{2,+}$, Patricia Fachini$^{1}$,  Lijuan Ruan$^{3}$, 
An Tai$^{4}$,\\
 Zhangbu Xu$^{1}$ and Haibin Zhang$^{1}$ (for the STAR Collaboration)\\\\
$^{1}$ {\it Brookhaven National Laboratory, NY, USA.}\\
$^{2}$ {\it Institute of Physics, Bhubaneswar, INDIA.}\\
$^{3}${\it Lawrence Berkeley National Laboratory, CA, USA. and } \\
$^{4}${\it University of California at Los Angeles, CA, USA.}\\
$^+$e-mail: dmishra@iopb.res.in}
\date{}
\maketitle

\begin{abstract}
The measurements of the transverse momentum spectra and the invariant mass 
distributions of $\Delta(1232)$ $\rightarrow \pi p$, $K^{*}(892)\rightarrow 
\pi K$ resonances in $d$+Au collisions at $\sqrt{s_{NN}}$ = 200 GeV using 
the STAR Time Projection Chamber (TPC) at RHIC are presented. 
The in-medium modification of the
$\Delta$ and $K^{*}$ mass and width has been studied as a function
of transverse momentum ($p_T$). The particle ratios $K^{*}/K$, $\Delta/p$ 
and the average transverse momentum ($\langle p_{T} \rangle$) as a function of
different collision centrality has been reported. The nuclear modification 
factors ($R_{CP}$ and $R_{dAu}$) of $\Delta$ and $K^{*}$ are discussed.

\end{abstract}

\section{Introduction}
In a relativistic heavy ion collision an extended hot and dense matter is 
formed. At such a state, it is expected that nuclear matter goes through 
a phase transition from a confined (hadronic) matter to a de-confined phase 
or quark-gluon plasma (QGP). The studies of hadronic resonance with extremely 
short lifetimes ($\sim$few fm/c) have the unique characteristics to probe 
the hadron production and the collision dynamics through their decays and 
re-generation. In the hot and dense matter, resonances and their hadronic 
decay daughters undergo re-scattering and re-generation, which affects 
various resonance properties such as yields, masses, widths and also in the 
modification in the observed momentum distributions of the resonances 
\cite{rho}. Resonances with higher $p_T$ have a greater chance to be 
detected than the ones with lower $p_T$. That means higher momentum 
resonances leave the medium very fast and decay 
outside the medium, hence their daughter particles interact less with 
the medium \cite{bleicher}. The effect of re-scattering of the decayed 
daughters from resonances can destroy part of the primordial resonance 
yields. On the 
other hand, hadrons can interact with each other inside the medium and can 
enhance the primordial resonance yields \cite{bleicher1}. Thus, measuring 
the resonance yields and their ratios with respect to the corresponding 
stable particles in heavy ion collisions, compared to the same in $p+p$ 
collisions, can provide the information about hadronization and the 
dynamics between the chemical and kinetic freeze-outs and possible 
medium effects.

In this report we present some results of the above mentioned study, through 
the production of $\Delta$ and $K^*$ resonance in $d$+Au collision data taken
by the STAR (Solenoidal Tracker At RHIC) experiment in 2002-2003 RHIC run. 
Comparing the results from $d$+Au collisions with the same from $p+p$ and 
Au+Au collisions will enables us to understand the in-medium effects.

\section{Analysis}

In the present analysis, the $\Delta$ and $K^*$ resonances signals were 
measured via their hadronic decay channels $\Delta^{++}\rightarrow$ $p\pi^+$,
$\overline{\Delta}^{--} \rightarrow \overline{p}\pi^-$, 
$K^{*0}\rightarrow K^+ \pi^-$, $\overline{K^{*0}}\rightarrow K^-\pi^+$
and $K^{*\pm}\rightarrow K_{S}^{0}\pi^{\pm}$ in $d$+Au collisions 
at $\sqrt{s_{NN}}$ = 200 GeV. The main tracking device
in the STAR experiment is the TPC \cite{tpc}, which
provides both the momentum information and the particle identification 
of the charged particles by measuring their ionization energy loss 
($dE/dx$) in the TPC.
The minimum bias trigger was defined by requiring at least one beam-rapidity 
neutron in the Zero Degree Calorimeter (ZDC) in the Au beam direction,
which is assigned the negative pseudorapidity ($\eta$) \cite{dautrig}. 
The centrality of the $d$+Au collisions was
determined by the charged particle multiplicity ($N_{ch}$) within a 
pseudorapidity window of -3.8 $< \eta <$ -2.8 was measured by the 
Forward TPC (FTPC) along the Au beam direction. The $d$+Au 
centrality definition consists of three event centrality classes; 0-20\%,
20-40\% and 40-100\% of the total $d$+Au inelastic cross section. 
In this analysis, about 11.7M and 15M $d$+Au minimum
bias events have been used for $\Delta$ and $K^*$ respectively. 
\begin{figure}[ht]
\epsfxsize=4.cm
\epsfysize=3.cm
\begin{minipage}{0.4\textwidth}
\epsfbox{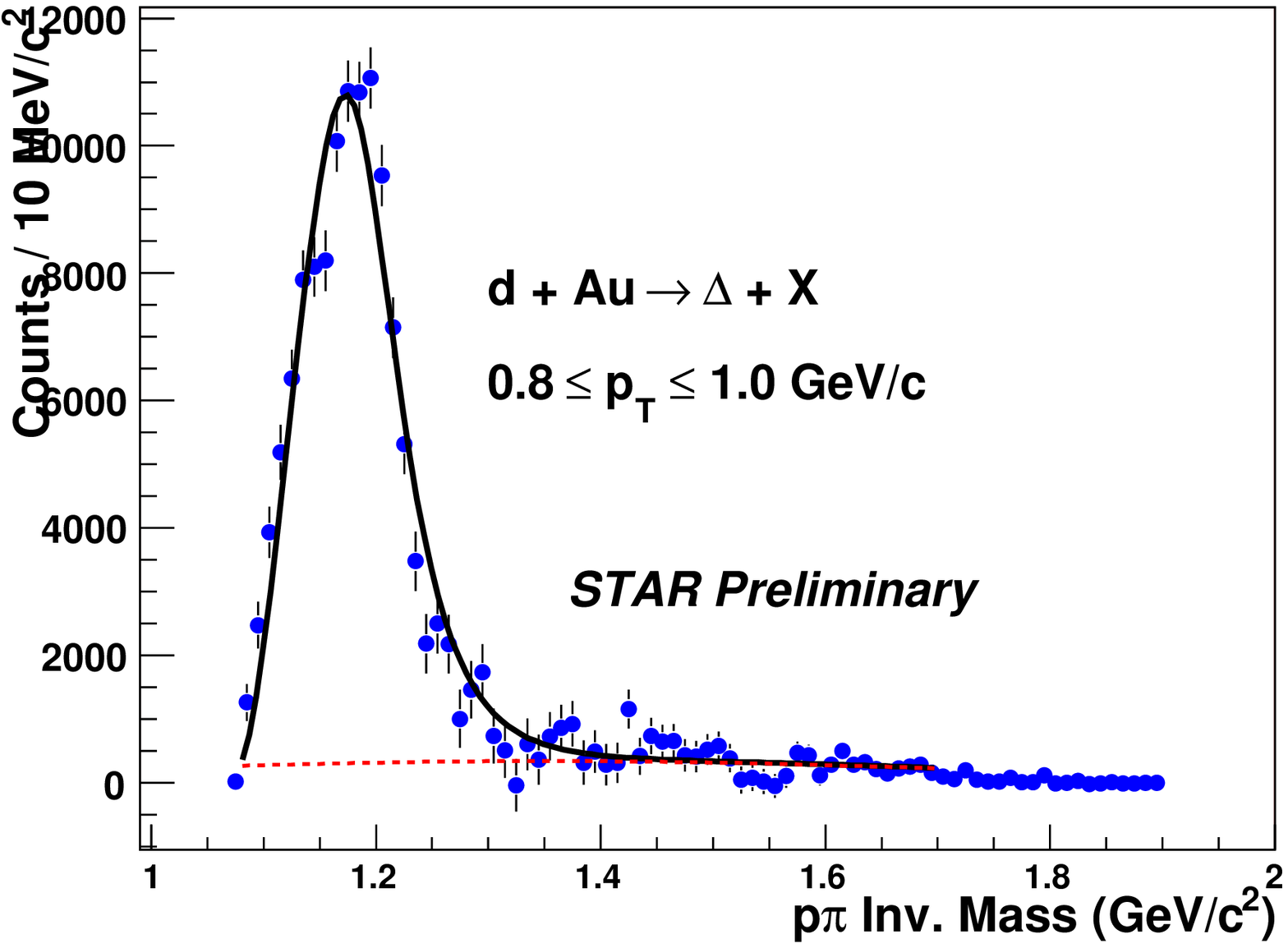}
\end{minipage}\
\epsfxsize=4.cm
\epsfysize=3.cm
\begin{minipage}{0.4\textwidth}
\epsfbox{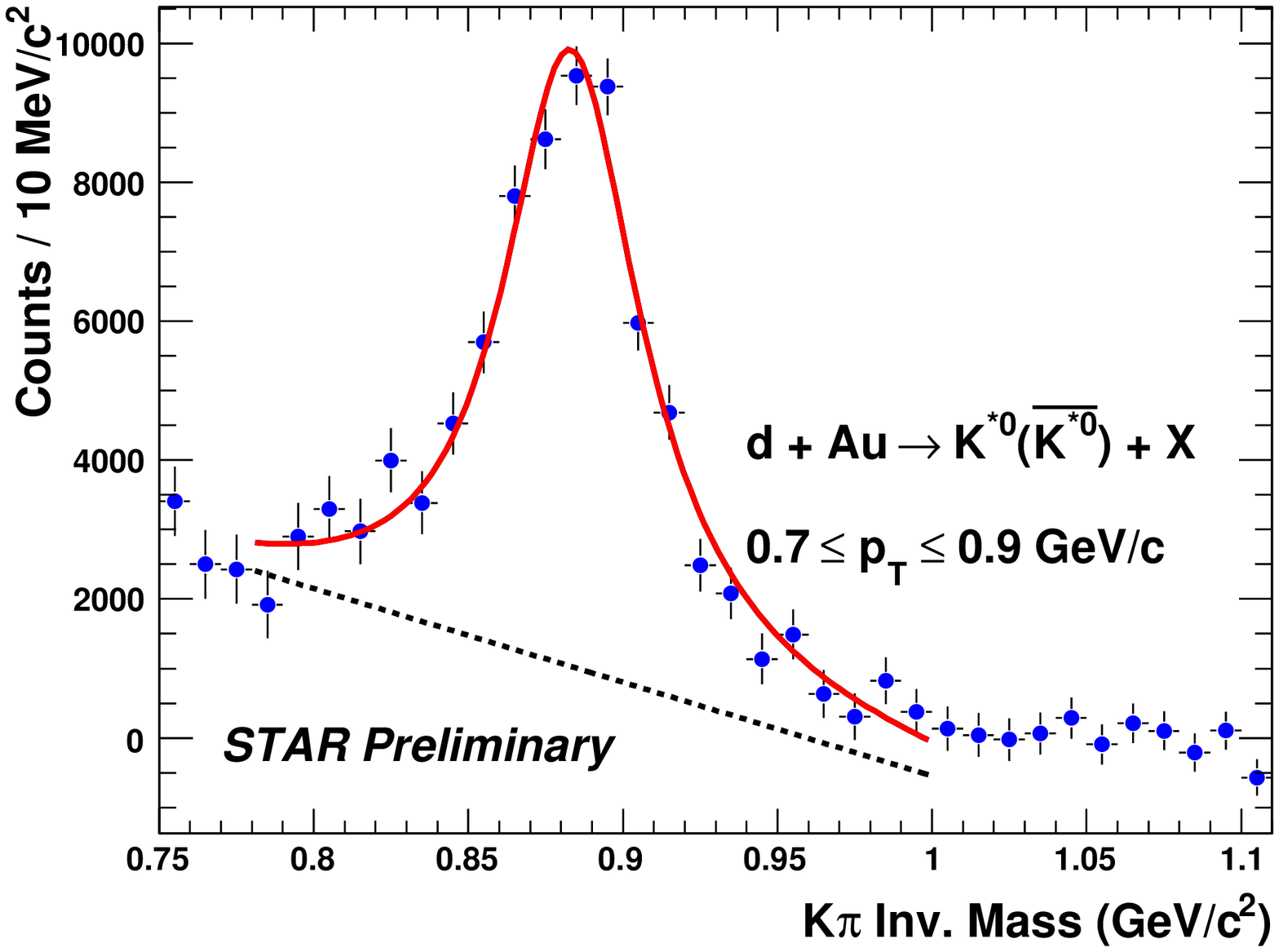}
\end{minipage}
\caption{(a) The $p\pi$ pair invariant mass spectrum. 
(b) The $K\pi$ pair invariant mass spectrum after 
mixed-event background subtraction fitted with BW $\times$ PS + background.}
\label{Fig: fig1}
\end{figure}
The $\Delta$ and $K^*$ 
invariant mass spectra are reconstructed using the event-mixing technique
\cite{eventmix}. 
Figure 1, shows the invariant mass spectra for
$\Delta$ and $K^*$, fitted with the $p$-wave Breit-Wigner function times the 
phase space factor (BW $\times$ PS) and a residual background function 
\cite{patricia,haibin}. For $\Delta$ case the residual background
is described by a Gaussian function, whereas for $K^*$ case it is 
described by a linear function.

\section{Results}

\begin{figure}[htp]
\begin{centering}
\hspace{2cm}
\epsfxsize=7cm
\epsfysize=3.cm
\epsfbox{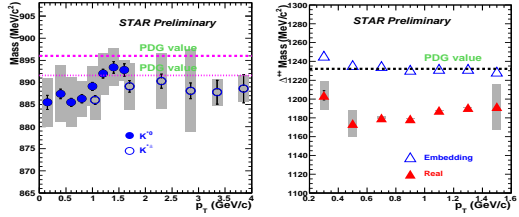}
\end{centering}
\caption{The resonance mass with their stat. and sys. uncertainties as a 
function of $p_T$, Left panel: for $K^{*}$, 
 Right panel: for $\Delta$ mass from real data (filled symbols) compared 
with the results obtained from Monte Carlo (MC) simulation (open symbols).}
\label{Fig:4.5}
\end{figure}
On the left panel of the Fig.2, one can see there is a 
mass shift for both $K^{*0}$ and $K^{*\pm}$ at lower $p_T$ bins. It is about 
10 MeV/c$^2$ less compared to the standard Particle Data Group (PDG) \cite{pdg}
value for $p_T<$ 0.9 GeV/c. The resonances produced with low $p_T$
spent more time inside the medium than the high $p_T$ resonances, resulting
in a modification in their masses. The high $p_T$ resonances leave the
medium very fast and don't get modified. On the right panel shows the 
$\Delta$ mass distribution for 0.2 $<p_T<$ 1.6 GeV/c.
Which shows the 
clear mass shift on the average up to about 50 MeV/c$^2$ observed. The
relatively smaller reconstructed masses compared to the standard value may be
attributed to the momentum loss from the re-scattering of decayed daughters.
\begin{figure}[hhtp]
\begin{centering}
\hspace{2cm}
\epsfxsize=7cm
\epsfysize=3.cm
\epsfbox{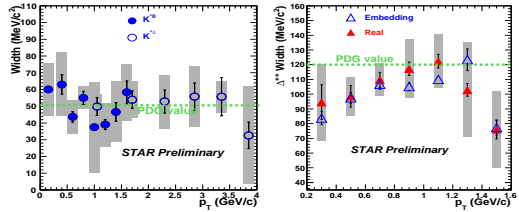}
\end{centering}
\caption{Left panel: the $K^{*}$ width with their stat. and sys. 
uncertainties as a function of $p_{T}$, Right panel: the $\Delta$ 
mass from real data (filled symbols) and the Monte Carlo (MC) 
simulation (open symbols) as a function of $p_{T}$.}
\label{Fig:4.5}
\end{figure}
In Fig.3, the width has been plotted as a function of $p_T$ for both $K^*$
and $\Delta$. On the left panels of the figure we don't see any difference 
between the width of $K^{*0}$ and $K^{*\pm}$ from their PDG value.
The right panel shows an increase in $\Delta$ width with increase in $p_T$, 
excepting the last $p_T$ bin, which is expected to be because of dynamical 
cut effects. 
\begin{figure}[ht]
\epsfxsize=4cm
\epsfysize=3.cm
\begin{minipage}{0.4\textwidth}
\epsfbox{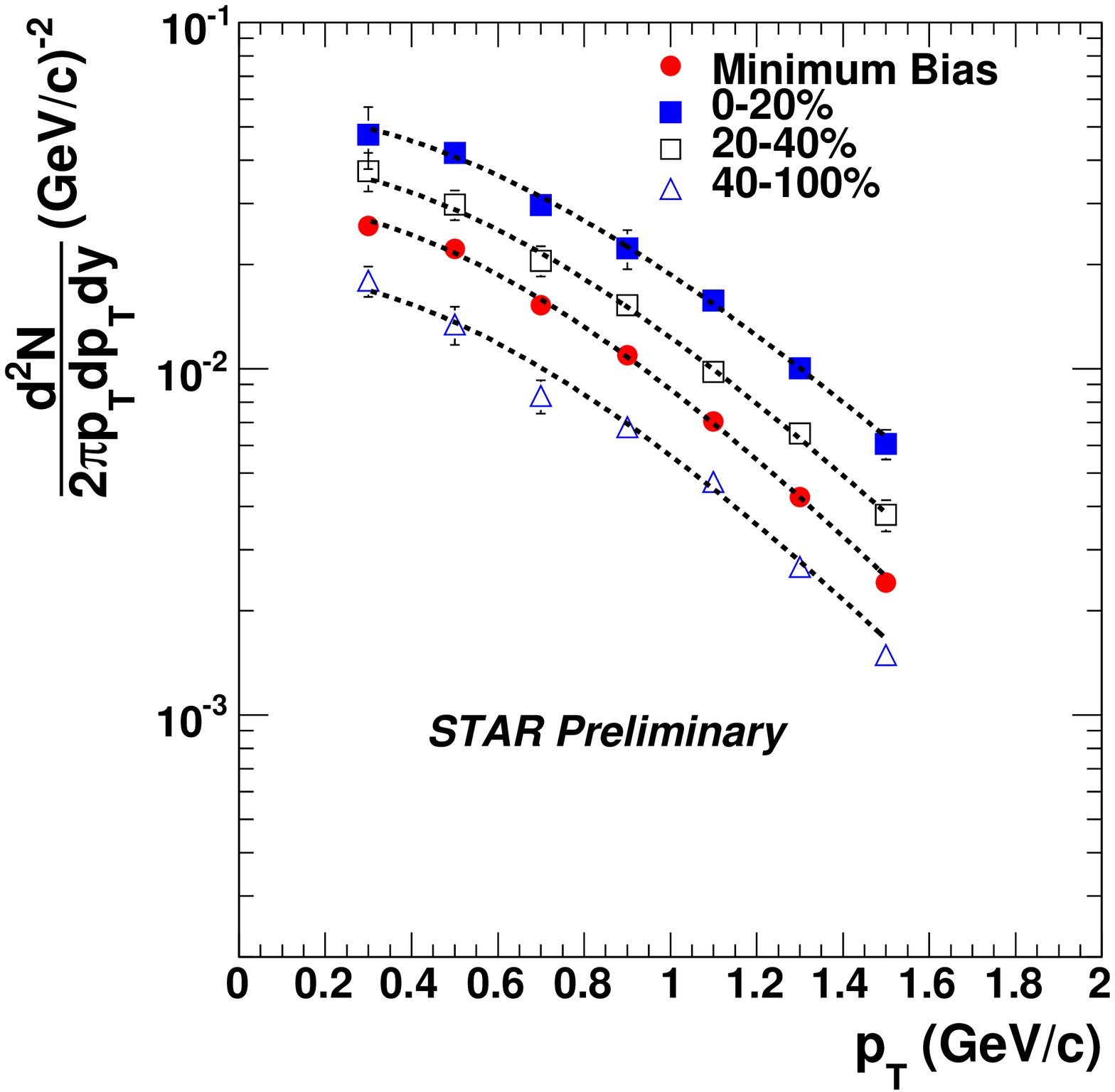}
\end{minipage}\
\epsfxsize=4cm
\epsfysize=3.cm
\begin{minipage}{0.4\textwidth}
\epsfbox{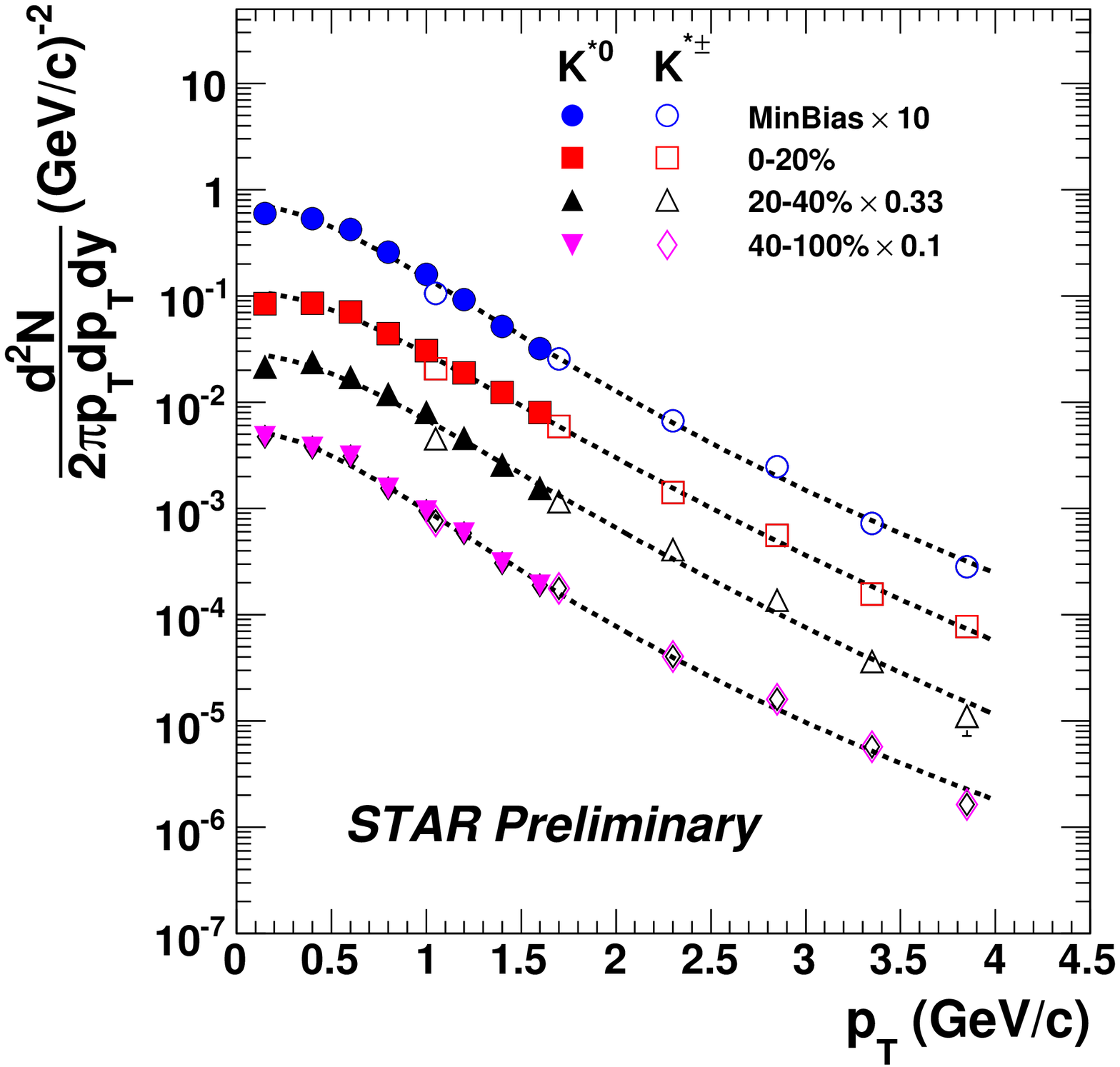}
\end{minipage}
\caption{The $p_T$ spectra for minimum bias as well as for different 
centralities, (a) for $\Delta$, fitted with exponential function and 
(b) for $K^*$, fitted with Levy function.}
\label{Fig: fig1}
\end{figure}
From the fit to the invariant mass spectra, the $\Delta$ and $K^*$ raw yield
has been extracted and corrected for efficiency and acceptance. Figure 4 
shows the corrected $p_T$ spectra for $\Delta$ and $K^{*}$ for minimum bias 
as well as for different collision centralities. From the fitting to the 
$p_T$ spectra one can get the resonance yield 
($dN/dy$) at mid-rapidity and the inverse slope parameter $T$ for $\Delta$
and $K^*$ for $d$+Au collisions \cite{haibin,dipak}. 
  \begin{figure}[ht]
\epsfxsize=4cm
\epsfysize=4cm
\begin{minipage}{0.3\textwidth}
\epsfbox{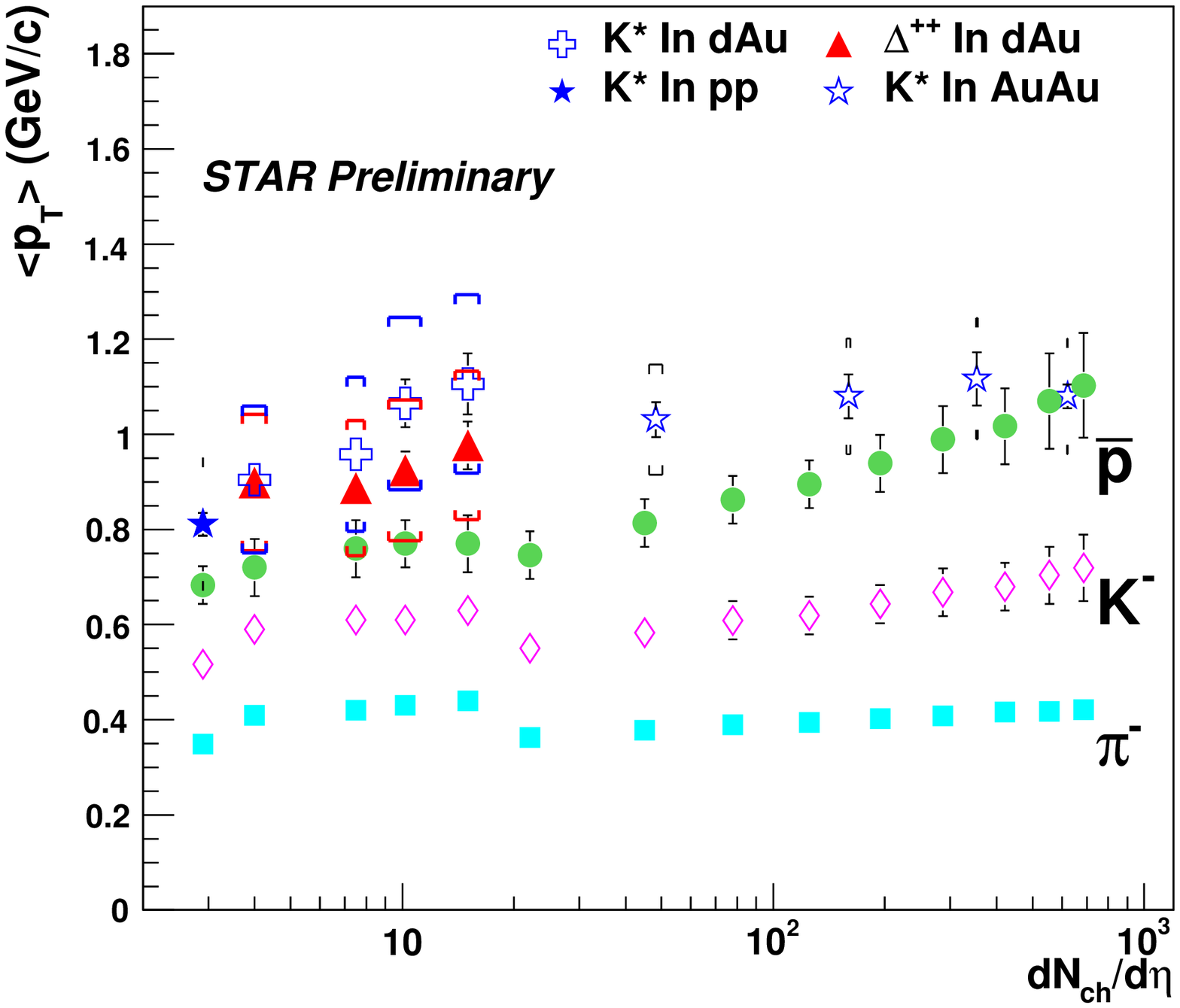}
\end{minipage}
\epsfxsize=4cm
\epsfysize=4cm
\begin{minipage}{0.3\textwidth}
\epsfbox{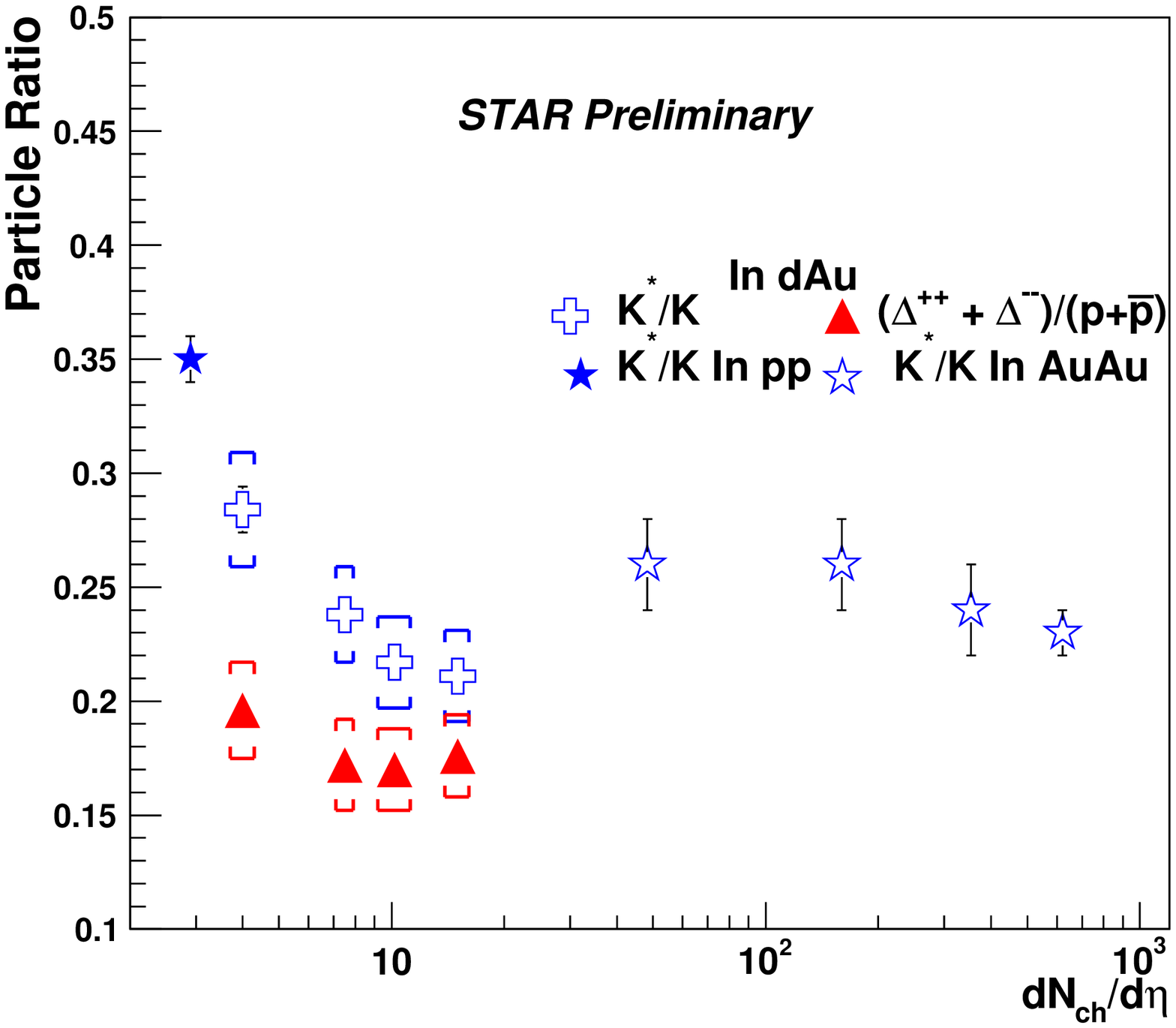}
\end{minipage}
\epsfxsize=4.5cm
\epsfysize=5cm
\begin{minipage}{0.3\textwidth}
\epsfbox{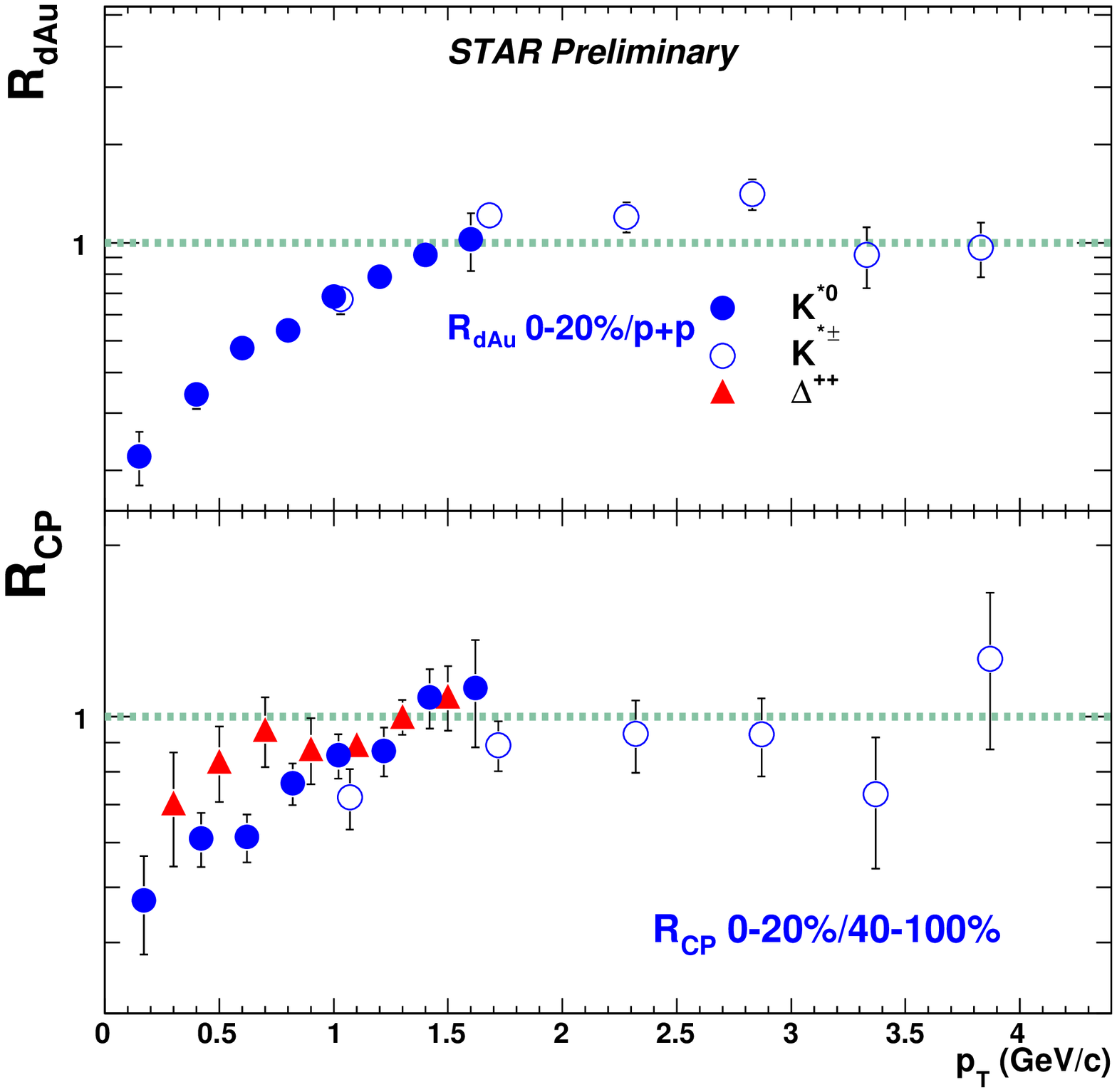}
\end{minipage}\
\caption{Left: The $\Delta$ and $K^*$ $\langle p_{T} \rangle$ as function of
  charged hadrons compared with $\pi^-$, $K^{*-}$, $\overline p$, middle: 
$K^*/K$ and $\Delta/p$ ratio as a function of charged hadrons, right: $\Delta$ 
and $K^*$ nuclear modification factor as a function of $p_T$.}
\label{Fig: fig1}
\end{figure}
As has been discussed earlier, the resonances having higher $p_T$ have greater
probability to be detected than the low $p_T$ ones, thus we expect higher 
mean transverse momentum ($\langle p_{T} \rangle$) values in heavy ion 
collisions than in elementary collisions, such as $p+p$ collisions 
\cite{bleicher1}. In the left panel of Fig.5, one can see, the 
$\langle p_{T} \rangle$ values for $K^*$
increase with centrality and higher than the same obtained from $p+p$ 
collisions at same center of mass energy, whereas there is no centrality
dependence observed in Au+Au collisions \cite{haibin}. In the same figure 
for the case of $\Delta$, there
is a slight increase in $\langle p_{T} \rangle$ as we go from peripheral 
to central collisions. The middle panel of the Fig.5, we can see that
the $K^*/K$ ratios in $d$+Au and Au+Au collisions are significantly smaller 
than the same in $p+p$ collisions. The $K^*/K$ ratio suppression may indicate
that between the chemical and kinetic freeze-out, $K^*$ signals are 
predominantly destroyed due to the re-scattering of daughter particles which
cannot be compensated by the re-generation effect. Where as for 
$(\Delta^{++}+\overline{\Delta}^{--})/(p+\overline p)$ ratios don't show
any centrality dependence in $d$+Au collisions. The right panel of the
Fig.5 shows the nuclear modification factor $R_{dAu}$ and $R_{CP}$ for 
$\Delta$ and $K^*$ as a function of $p_T$. The $R_{CP}$ values are less 
than unity at low $p_T$ region. But $R_{dAu}$ and $R_{CP}$ close to unity 
at $p_T>$ 1.5 GeV/c for both $\Delta$ and $K^*$. The lower value of $R_{CP}$
for low $p_T$ region seem to be a result of re-scattering of daughter 
particles inside the medium.

\section{Conclusions}
The preliminary results on $\Delta$ and $K^*$ resonances measured
using the TPC in STAR experiment at mid-rapidity in $d$+Au collisions 
at $\sqrt{s_{NN}}$ are reported. The downward mass shift ($\sim$ 10 MeV) 
observed at lower $p_T$ bins are observed for $K^*$ and there is a 
clear mass shift (up to $\sim$ 50 MeV) observed for $\Delta$ resonance 
over all $p_T$ bins. The observed $K^*/K$ ratios in $d$+Au collisions are
significantly smaller than that for $p+p$ collisions. On the other hand 
$\Delta/p$ ratios in $d$Au collisions are independent of centrality. 
The $\langle p_{T} \rangle$ for $K^*$ in $d$Au collisions is higher
than the same obtained from $p+p$ collisions, in agreement with
$p_T$-dependence of daughter particles' re-scattering effect. The 
nuclear modification factor is close to unity for both $\Delta$ and 
$K^*$ resonaces at $p_T>$ 1.5 GeV/c.

\end{document}